\documentstyle[12pt]{article}

\begin{document}
\begin{titlepage}
\title { Finite-size effects in multiquark droplets; An anatomical
study}
\author {{V. S. Uma Maheswari }\\
{\it Variable Energy Cyclotron Centre}\\
{\it 1/AF Bidhan Nagar, Calcutta - 700064, India.}}
\maketitle
\begin{abstract}
{ Strutinsky's averaging(SA) method is applied to multiquark droplets
to systematically extract the smooth part of the exact quantal energy
and thereby the shell correction energies. It is shown within the
bag model that the semi-phenomenological density of states expression
given upto curvature order is almost equivalent to the SA method.
A comparative study of the bag model and the relativistic harmonic
oscillator potential for quarks is done to investigate the quark
mass dependence of the finite-size effects. It is found that there
is an important difference between these two cases, which may be
related to the presence/non-presence of the net spin-orbit effect.}
\end{abstract}
\end{titlepage}
\newpage
\noindent {\large {\bf I.\ Introduction}}
\vskip 1.5 true cm

The most interesting prediction in recent times is the Witten's
conjecture\cite{witten}
that the strange quark matter(SQM) might be the absolute ground-state of
hadronic matter. Such a possibility has very significant astrophysical
consequences\cite{astro}. In addition, probable formation and detection of small lumps
of SQM in relativistic heavy-ion collisions is being promoted as an
unambiguous signature of quark-gluon plasma\cite{greiner}.
In this context, stability of strangelets with baryon number $A\le 100$
is of immense interest. In the determining the stability criteria of
these small strangelets the finite-size effects such as surface and curvature
play a significant role.

Farhi and Jaffe\cite{farhi} were the first to consider the surface effects.
Later on, Mardor and Svetitsky\cite{mardor} suggested that curvature effects
are also important.
Recently, Madsen\cite{madsen} has given a semi-phenomenological density of
states(DOS) expression including terms upto curvature order for a system
of quarks with mass $m_q$.
He has shown within the bag model that this DOS expression reproduces
quite well the exact shell model calculations irrespective of the quark
mass $m_q$. (We hereafter refer to this model as quark liquid drop(QLD)
model.)
Further, an intriguing aspect is that the finite-size contributions in
the QLD model do not seem to show a converging trend.
More specifically, in the case of massless quarks the surface energy
is zero, whereas the curvature
contribution is significant. And in the case of massive quarks,
the surface energy contribution is non-zero while the curvature energy
contribution can be zero or positive or negative depending on the value
of quark mass $m_q$. ( These features are clearly illustrated in
Fig.1 of Ref.\cite{madsen} ).
Consequently, the question of the importance of still
higher-order terms such as Gauss curvature naturally arises. In other words,
is the agreement between the QLD and shell model calculations accidental?

This question can be answered rather quantitatively with the help of the
famous Strutinsky's averaging(SA) method\cite{strutinsky}.
In the context of nuclear physics,
the  SA method and the Wigner-Kirkwood(WK) expansion\cite{wigner} have been
shown\cite{bhadhuri} analytically to be equivalent. Therefore,
by comparing the QLD model and the SA method we can determine
 the importance/non-importance of higher-order terms
in the semi-phenomenological DOS expression of Madsen .
In addition, the QLD model suggests that the finite-size effects
depend on the value of the quark mass. As this can have very significant
consequences in the study of quark-hadron phase transition\cite{mardor},
 it would be
interesting to know whether this finding is of general nature or is
model-dependent.  In this work, we mainly focus upon these two aspects
of multiquark droplets. For this purpose, we consider the two potentials;
the infinite square well(ISqW), namely a spherical cavity and the
relativistic harmonic oscillator(RHO). The beauty of these two potentials
is that both mimic quite adequately the asymptotic freedom and
confinement properties of QCD. Moreover, they are analytically
solvable.

In section II, we briefly discuss the nature of the shell
structure in both these potentials. The Strutinsky's averaging method
appropriate for quark systems is presented in section III and the results
obtained are discussed in section IV. Finally, we summarize our findings in
section V.
 \vskip 1.0 true cm

\noindent {\large {\bf II.\ Confining potential, quark mass and shell structure}}
\vskip 1.0 true cm

In regard to the shell structure in the ISqW and RHO potentials, at the
outset the following observations can be made.

\begin{itemize}
\item
In the case of the spherical cavity, the eigenvalues obtained
exhibit splitting of levels such as $(p_{3/2}, p_{1/2})$, $(d_{5/2}, d_{3/2})$
$\cdots$ due to the spin-orbit coupling. On the otherhand, in the harmonic
oscillator the eigenvalues are $j-$independent and hence no net spin-orbit
effect on the energy spectrum. Thus, as expected, the shell structure
depends on the nature of the confining potential.
\item
The other interesting observation is that the energy gap $\Delta_j$ between
the $l=j+1/2$ and $l=j-1/2$ levels decreases as the value of quark mass $m_q$
increases. This is illustrated in Fig.1 where we have plotted the eigenvalues
obtained by the bag boundary condition\cite{bag} taking $m_q =0$, 150, 450 MeV.
The zero of the scale is taken at the $0s_{1/2}$ level of massless case and
therefore, what is plotted is $(\omega(\kappa) - 2.0428)$ in units
of $\hbar c/R$. ( While solving the boundary condition for massive quarks
we have taken $R=1$ fm).
The decrease in $\Delta_j$ can be clearly noted in the $0p$ and $0d$ shells.
In addition, due to the dependence of $\Delta_j$ on $m_q$, rearrangement
of levels occur at higher $j$ values as compared to the massless quark
spectrum. The degree of rearrangement ofcourse depends upon the magnitude
of $m_q$.
In a recent work\cite{sharma}, it was found within the relativistic mean field theory
of nuclei that the spin-orbit term is inversely proportional to the
density dependent nucleon effective mass.
In analogy, we may expect $\Delta_j$ to decrease as the quark
mass increases. Results displayed in Fig.1 seem to illustrate this effect.
On the otherhand, in the case of RHO potential, the shell structure remains
more or less the same expect for a relative shift in energy as $m_q$
is varied.
\end{itemize}

Thus, already one can see that the nature of the confining potential
and the quark mass have  definite roles to play in determining the
shell structure and thereby the stability of small multiquark droplets.
In the following section, we shall outline the SA method and apply it
to a system of quarks.
\vskip 1.0 true cm

\noindent {\large {\bf III.\ Strutinsky's averaging method for quark systems}}
\vskip 1.0 true cm

The basic concept of the SA method\cite{strutinsky} is that the total
energy $E$ of a given quantal system can be separated into a smooth part
${\bar E}$ and an oscillating part $\Delta E $, i.e. $E= {\bar E} + \Delta E$.
For a multiquark system consisting of more than one flavor, $\Delta E$
pertaining to each flavor needs to be calculated separately.
For the sake of simplicity, we consider here a system of one kind of
quarks with mass $m_q$ confined either in a spherical cavity  or in a
harmonic oscillator potential. Then the exact single particle DOS is given by
\begin{equation}
g ( \epsilon ) = g \sum_{i} { \delta (\epsilon - \epsilon_i ) },
\end{equation}
where $g$ is the degeneracy factor and $\epsilon_i$ are the eigenvalues
of the potential under consideration. The total energy of the quark
system is then given as,
\begin{equation}
E= g \sum_m \epsilon_m \quad ; \quad m\ = \ {\rm occupied\ levels}.
\end{equation}
Now, to evaluate the smooth part of the total energy $E$, Strutinsky
proposed a numerical averaging of the single particle spectrum
$g(\epsilon )$ by Gaussian smoothing funtions centered around $\epsilon_i$
and taken over a certain energy range $\gamma$. Thus, the smoothed level
density is given as
\begin{equation}
{\bar g}(\epsilon ) = {g\over \gamma \sqrt {\pi}} \sum_{m=1}^{M} exp (-U_m^2)
 { \sum_{i=1}^p C_i H_i ( U_m) },
 \end{equation}
where $U_m = (\epsilon - \epsilon_m )/\gamma $, $M$ is the number of levels
taken into consideration and $p=6$\cite{nix}.
The coefficients $C_i =(-1)^{i/2}/ [ 2^i (i/2)! ]$ for even $i$ values, and
$C_i =0$ for odd $i$ values. $H_i (U_m)$ are the Hermite polynomials.
The average number of particles ${\bar N}$ and ${\bar E}$ are then
defined as,
\begin{eqnarray}
{\bar N} &=& \int_{\infty}^{\bar \lambda} {\bar g}( \epsilon ) d\epsilon
, \nonumber \\
{\bar E} &=& \int_{\infty}^{\bar \lambda} \epsilon {\bar g}( \epsilon ) d\epsilon.
\end{eqnarray}
Having obtained the exact value $E$ and the smooth part ${\bar E}$,
we can determine the shell correction energies using the definition
$ \Delta E = E - {\bar E}$. In the following, we apply this method
to the RHO and ISqW potentials.
\vskip 1.0 true cm

\centerline {\bf A.\ Relativistic harmonic oscillator potential }

\vskip 1.0 true cm
Considering the quarks with current quark mass $m_q$ to be confined by
a scalar plus vector harmonic confinement of the type
${1\over 2} (1+\gamma_0 )( m_q+ C_2 r^2 )$, the corresponding Dirac equation
can be solved analytially\cite{smith}. And the eigenvalues are determined
by the equation
\begin{equation}
(\epsilon_N -m_q)(\epsilon_N^2-m_q^2 ) = 4C_2 (N+{3\over 2})^2,
\end{equation}
with $N=2n+l$. In the non-relativistic limit,
$\epsilon_N= (N+3/2)\sqrt{(2C_2/m_q)} + m_q$ and in the limit of small
mass, $\epsilon_N \simeq {[2\sqrt {C_2} (N+3/2)]}^{2/3} + m_q/3 $.
In analogy with the non-relativistic case, we take the potential
strength parameter $C_2=C_{20}A^{-2/3}$, where $A=N_q/3$ and $N_q$ is the
number of quarks. The quantity $C_{20}$ can be related to the ground-state
energy $M_b$ of a $3-q$ system as $C_{20} = (1/9){(M_b /3)}^3$.
In the case of massless quarks, $M_b$ = 1085.5 MeV and $C_{20}=(173.95)^3$
MeV$^3$; and for massive quarks, $M_b$ = 1672.45 MeV and $C_{20}=(268.01)^3$
MeV$^3$.

The next important step is to choose an appropriate value of $\gamma$, and
check that the smoothed part does not depend upon $\gamma$ value. This can
be done by using the so-called plateau condition\cite{brack}.
In our study, we
choose the $\gamma-$ parameter to be $\gamma = \gamma_0 C_{20}^{1/3} A^{-1/3}$.
Then, the value of $\Delta E$ corresponding to the plateau region pertaining
to each value of $m_q$ and $A$ is taken to be the `physical' shell correction
energy which then determines the value of ${\bar E}$.
Further, it was found that the prominence
of the plateau and its length depends upon the number of quarks $N_q$.
As $A$ decreases, it becomes somewhat tricky to fix the value of $\Delta E$.
The indeterminacy is however small.

It would be interesting to compare the so-determined values of ${\bar E}$ 
with those
obtained using a Wigner-Kirkwood expansion and study the importance
of higher-order finite-size effects such as Gauss curvature.
The expressions for the quark number and the total energy using
the WK expansion can be obtained analytically\cite{vinas}.
We state below the final expressions so-obtained upto $O(\hbar^2)$.
\begin{eqnarray}
N_q  &=& m_q^3 {\left ( {2m_q\over C_2} \right )}^{3/2} (1+x)^{3/2}x^3
       -{3m_q\over 4} {\left ( {2m_q\over C_2} \right )}^{1/2} (1+x)^{1/2}x,
       \nonumber \\
E_{WK} &=& {64\over 1155}\ m_q^4 {\left ( {2m_q\over C_2} \right )}^{3/2}
            (1+x)^{1/2} \cdot \nonumber \\
&\quad& \left [ {945\over 32} x^5 +{2905\over 64}x^4 + {1135\over 64}x^3 +
{3\over 8}x^2 -{1\over 2}x + 1 - (1+x)^{-1/2} \right ] \nonumber \\
& \quad & - {2\over 5} m_q^2 {\left ( {2m_q\over C_2} \right )}^{1/2}(1+x)^{1/2}\
\left [ {9\over 4}x^2 +{11\over 8}x +1 - (1+x)^{-1/2} \right ],
\end{eqnarray}
where $x=(\eta-m_q)/m_q$ and $\eta$ is fixed by the number equation.
Before, we present our results in this case, we would like to discuss the SA
method in the case of ISqW potential.
\vskip 1.0 true cm

\centerline {\bf B.\ Infinite square well potential }
\vskip 1.0 true cm

Considering $N_q$ number of quarks with mass $m_q$ to be confined in a
spherical cavity of size $R$, the boundary condition\cite{bag} to be solved for
determining the eigenvalues is
\begin{equation}
j_l (\omega_{\mu}) = -sgn(\kappa ){\omega_{\mu }\over {\omega +\mu}}
j_{\bar l} (\omega_{\mu}),
\end{equation}
where $\omega^2=\omega_{\mu}^2 + \mu^2 $ with $\mu=m_q R$, and
${\bar l} = l-sgn(\kappa )$. Then, the total energy of the system
using the bag model picture is,
\begin{equation}
E = {\hbar c\over R}\sum_i \omega_i + {4\over 3}\pi R^3 B,
\end{equation}
where $B$ is the bag energy density. The equilibrium radius of the system
is then determined by the saturation condition, ${\partial E/\partial R}
\mid_{R_o} =0$.

In this case, the single particle DOS $g(\omega )= \sum_i (\omega-\omega_i)$
is smoothed as in Eq.(3) and is given as,
\begin{equation}
{\bar g}(\omega ) = {g\over \gamma_{\omega} \sqrt {\pi}} \sum_{m=1}^{M}
 exp (-U_m^2)  { \sum_{i=1}^p C_i H_i ( U_m) },
 \end{equation}
with $U_m = (\omega - \omega_m)/\gamma_{\omega}$.
Then the smoothed energy ${\bar E}$ can be obtained using the equation,
\begin{equation}
{\bar E} = {\hbar c\over R}{\bar E_{\omega}}(R) + {4\over 3}\pi R^3 B,
\end{equation}
where $ {\bar E_{\omega }} = \int_{\infty}^{\lambda_{\omega}}
\omega {\bar g(\omega )} d\omega $ and $\lambda_{\omega }$ is determined by
the number equation.
Here, the $\gamma_{\omega }$ is chosen as;
$\gamma_{\omega} = \gamma_{0} A^{-1/3}$.

With this parametrisation, we have checked the plateau condition in this
case also for each value of $m_q$ and $A$ while calculating ${\bar E}$.
\vskip 1.0 true cm

\noindent {\large {\bf IV.\ Results and discussions }}
\vskip 1.0 true cm

Having demonstrated a reliable way of extracting the smooth part from the
exact quantal energy, we shall presently compare this SA method with the
WK one in the case of RHO potential, and the QLD model in the
case of ISqW potential.

In Table I, we have presented the energies obtained in the case of RHO
potential for four values of baryon number $A$ using the SA method
[Eqs.(3-5)] and the WK method[Eq.(6)] and then are compared with exact ones[Eq.(2)].
The difference between ${\bar E}$ and $E_{WK}$ is about 10 MeV for massless
quarks and about 17 MeV for the massive ones. This may be attributed to the
presence of higher-order $[O(\hbar^4)]$ correction terms. Notwithstanding
this, we might say that the agreement is quite good. This then establishes
the goodness of the WK expansion for further study of the quark mass
dependence of surface and curvature energies.

In Table II, the results obtained in the case of the ISqW potential is given.
We have taken $m_q=1$ MeV and $B^{1/4}$=145 MeV.
One can immediately see that $E_{QLD}$ agrees quite well with that of
the SA method. Similar degree of agreement is also obtained for more
massive quarks. This then demonstrates that the DOS
expression upto curvature order of Madsen is quite adequate.

 Consequently, an important question arises; Does the finite-size effects
greatly depend on the mass of the quark as suggested by the QLD model?
To answer this, we need to understand the following aspects of the
liquid drop model(LDM) expansion of energy\cite{brack1,ls}.
The surface and curvature energy coefficients contributes respectively to orders
of $A^{2/3}$ and $A^{1/3}$ in the LDM expansion of total energy:
\begin{equation}
E = a_v A + a_s A^{2/3} + a_{\rm cv} A^{1/3} + \cdots.
\end{equation}
To these orders, there is a part coming purely from the surface and curvature
terms of the DOS expression. But, there is also some extra contributions
arising out of the finite size effect on the Fermi momentum $k_F$.
( This can be noted from Eq.(4) of Ref.\cite{ls} ). Because of this,
the volume term in the DOS expression contributes towards $O(A^{2/3})$,
$O(A^{1/3})$ $\cdots$, and similarly the surface term contributes
towards $O(A^{1/3})$ and $O(A^{0})$, and so on. Therefore, the vanishing
of curvature term in the DOS expression does not necessarily mean that
curvature energy coefficient is zero, but ofcourse its value is expected
to be small.
In view of this, we would like to extract the effective surface and
curvature coefficients by making a least-squares fit to the total
energies $E_{QLD}$ calulated using three values of $m_q$.
To do so, we used the QLD model with $B^{1/4}=145$ MeV and have
obtained energies for 101 number of droplets with baryon number
$A$ such that $6 \le A \le 1000 $. The results so-obtained for
the energy coefficients are given in Table III for
$m_q = 0, \ 150 \ \& \ 450$ MeV.

It can be seen that the volume $a_v$ coefficients obtained here agree
with the bulk limits shown in Fig.1 of Ref.\cite{madsen} for
all the three values of $m_q$. Further, in the case of massless quarks
the surface energy coefficient $a_s$ is indeed zero,
and $a_{\rm cv}$ is about
220 MeV. As the quark mass is increased to 150 MeV, the curvature
coefficient has decreased as compared to the massless case illustrating
the fact that the curvature term in the DOS expression is nearly zero
for $m_q=150$ MeV. Similarly, the value of $a_s$ is now non-zero due to
the presence of a surface term in the DOS expression.
With further increase in $m_q$, the value of $a_s$ increases, while
$a_{\rm cv}$ remains more or less the same.
Hence, although the curvature contribution in the DOS expression becomes
negative as $m_q$ increases, the effective curvature energy remains
positive. Further, except for the massless case there is a converging trend
in the LDM expansion of energy. It must be said here that the estimates
obtained for higher-order coefficients such as Gauss curvature $(O(A^{0}))$ are
not reliable as such terms were not included in the DOS expression.
We have given them just to show that the three principal terms $a_v$, $a_s$ and
$a_{\rm cv}$ stabilises with respect to the number of parameters in the fit.
Thus, there is a systematic dependence of surface and curvature energies
on the mass of the quark. It must be stressed that the values obtained here
are dependent on the bag constant value.
Now, we are curious to see how far these findings are true in the
case of RHO potential.

For this purpose, we repeated the same exercise taking the WK expansion
of number and energy(Eq.(6)) for the same set of $A$ and $m_q$. Here
also we have checked the stability of the three energy coefficients with
respect to the number of parameters in the fit. The results so obtained
are given in Table IV. It can be seen that the surface
coefficient is independent of the mass of the quark.
And the curvature coefficient increases from 150 MeV to 220 MeV as $m_q$
varies from 0 to 150 MeV. With further increase in $m_q$, there is a
slight decrease in $a_{\rm cv}$. In the case of massless quarks, it is
somewhat disturbing to find the volume coefficient $a_v$ less than 930 MeV.
This can 
be rectified, as done in the case of bag models in choosing $B$,
by appropriately choosing the potential parameter $C_{20}$
so that the $ud$ matter is unbound against nuclear matter.
But, we feel our findings regarding the quark mass dependence of
the finite-size effects shall remain unaffected as the surface
and curvature terms are dependent only on the difference between
the total energy and the bulk value.

Further,
the surface energy coefficient remaining zero irrespective of the value
of the quark mass needs more critical examination as it can have significant
consequences for the phase transition studies. It may be recalled here that
in the case of ISqW potential, the shell structure depends on the quark mass
through the spin-orbit effect; whereas, in the case of RHO potential there
is no such effect. Is this the underlying reason for the weak dependence
of $a_s$ and $a_{\rm cv}$ on $m_q$ in the case of RHO potential? May be.

Thus, we have shown that the shell structure as well as the quark mass
dependence of the finite-size effects are dependent on the nature of
the confining potential.
\vskip 1.0 true cm

\noindent {\large {\bf VI.\ Summary}}
\vskip 1.0 true cm

In summary, we have applied the Strutinsky's averaging(SA) method
to multiquark droplets to systematically extract the smooth part of
the exact quantal energy, and thereby the shell correction energies.
It is shown in the case of bag model picture that the DOS expression
given upto curvature order reproduces quite well the smoothed
energies obtained by the SA method.
Similarly, we found in the case of relativistic harmonic oscillator(RHO)
potential, the Wigner-Kirkwood expansion with terms upto $O(\hbar^2)$
 reproduces well the smoothed values of energy.

Having established the goodness of the asymptotic expansions in
both the cases, we then made a comparative study of the two
potentials, namely the spherical cavity and the relativistic
harmonic potential, in regard to the quark mass dependence of the
finite-size effects.

It was found in the case of the RHO potential that in contrast to the
bag model picture, the surface and curvature coefficients are
weakly dependent on the quark mass $m_q$. Further, the surface energy
contribution is almost zero irrespective of the value of $m_q$.
These differences may be traced back to the difference in the shell
structure between these two potentials due to the presence/non-presence
of net spin-orbit effect.
\vskip 1.0 true cm

\noindent {\bf Acknowledgements:} Useful discussions with 
Professor J. Madsen and Dr. M.G. Mustafa are gratefully
acknowledged.

\newpage
\centerline {\bf TABLE CAPTIONS }
\vskip 1.0 true cm

\noindent {\bf Table I \ : \ } Relativistic harmonic oscillator potential.
Total energies calulated using the Wigner-Kirkwood(WK) expansion and
the Strutinsky's averaging(SA) method are compared with the exact
quantal ones for four values of baryon number $A$. The first set
corresponds to the massless quarks, and the next one corresponds
to quark mass $m_q=150$ MeV.
\vskip 1.0 true cm

\noindent {\bf Table II \ : \ } Infinite square well potential.
Total energies calulated using the quark liquid drop(QLD) and
the Strutinsky's averaging(SA) method are compared with the exact
quantal ones for four values of baryon number $A$ taking 
the quark mass $m_q$=1 MeV.
\vskip 1.0 true cm

\noindent {\bf Table III \ : \ } Infinite square well potential.
 Values of the energy coefficients, volume $a_v$,
surface $a_s$ and curvature $a_{\rm cv}$, obtained by making a least-squares
fit to the total energies calculated for 101 values of the baryon number $A$
in the range 6-1000 using the quark liquid model. The three sets corresponds
to quark mass $m_q$ = 0, 150, $\&$ 450 MeV.
\vskip 1.0 true cm

\noindent {\bf Table IV \ : \ } Relativistic harmonic oscillator potential.
Values of the energy coefficients, volume $a_v$,
surface $a_s$ and curvature $a_{\rm cv}$, obtained by making a least-squares
fit to the total energies calculated for 101 values of the baryon number $A$
in the range 6-1000 using the Wigner-Kirkwood expansion method.
The three sets corresponds
to quark mass $m_q$ = 0, 150, $\&$ 450 MeV.
\newpage
\centerline {\bf FIGURE CAPTIONS }
\vskip 1.0 true cm

\noindent {\bf FIG.1 \ : \ } Single particle levels obtained by solving
the bag boundary condition are shown relative to the $0s_{1/2}$ level
of the massless quark case in units of $\hbar c/R$. Index = 1, 2, $\&$ 3
corresponds to the quark mass $m_q$=0, 150 and 450 MeV respectively.
\vskip 1.0 true cm

\newpage
\centerline {\bf Table I }
\vspace {0.3in}
\begin{center}
\begin{tabular}{|c|c|c|c|}
\hline
\multicolumn{1}{|c|}{ $A$ }&
\multicolumn{1}{|c|}{ $E_{ex}$ }&
\multicolumn{1}{|c|}{ $E_{SA}$ }&
\multicolumn{1}{|c|}{ $E_{WK}$ }\\
\hline
6 & 5556.9 & 5466.2 & 5476.8 \\
20 & 17597.3 & 17713.5 & 17723.6 \\
40 & 34979.6 & 35126.0 & 35134.1 \\
70 & 61013.4 & 61189.4 & 61196.4 \\
\hline
6 & 9794.6 & 9653.1 & 9670.0 \\
20 & 31275.6 & 31449.0 & 31460.7 \\
40 & 62239.8 &62456.6  & 62467.2 \\
70 & 108629.3 & 108883.9 & 108899.7 \\
\hline
\end{tabular}
\end{center}
\vskip 1.0 true cm
\centerline {\bf Table II }
\vspace {0.3in}
\begin{center}
\begin{tabular}{|c|c|c|c|}
\hline
\multicolumn{1}{|c|}{ $A$ }&
\multicolumn{1}{|c|}{ $E_{ex}$ }&
\multicolumn{1}{|c|}{ $E_{SA}$ }&
\multicolumn{1}{|c|}{ $E_{QLD}$ }\\
\hline
6 &  6924.4  & 6948.9  &  6965.4  \\
20 & 22580.2 & 22429.0 & 22437.7  \\
40 & 44581.0 & 44406.5 & 44413.6 \\
70 & 77231.6 & 77291.1 & 77296.4  \\
\hline
\end{tabular}
\end{center}
\newpage
\centerline {\bf Table III }
\vspace {0.3in}
\begin{center}
\begin{tabular}{|c|c|c|c|c|c|}
\hline
\multicolumn{1}{|c|}{ $\#$ Param. }&
\multicolumn{1}{|c|}{ $a_v$ }&
\multicolumn{1}{|c|}{ $a_s$ }&
\multicolumn{1}{|c|}{ $a_{\rm cv}$ }&
\multicolumn{1}{|c|}{ $a_x$ }&
\multicolumn{1}{|c|}{ $a_y$ }\\
\hline
3 & 1090.3 & 1.258 & 218.0 & & \\
4 & 1090.4 & $-$0.5139 & 225.3 & $-$8.45 & \\
5 & 1090.4 & $-$0.0296 & 222.3 & $-$0.838 & $-$6.39\\
\hline
3 & 1212.4 & 265.0  & 91.8 & & \\
4 & 1211.8 & 273.7  & 55.9 & 41.4 & \\
5 & 1211.9 & 272.8  & 61.4 & 27.7 & 11.5\\
\hline
3 & 1837.8 & 303.4 & 78.0 & & \\
4 & 1837.5 & 307.5 & 61.0 & 19.6 & \\
5 & 1837.5 & 306.8 & 65.3 &  8.78& 9.10 \\
\hline
\end{tabular}
\end{center}
\newpage
\centerline {\bf Table IV }
\vspace {0.3in}
\begin{center}
\begin{tabular}{|c|c|c|c|c|c|}
\hline
\multicolumn{1}{|c|}{ $\#$ Param. }&
\multicolumn{1}{|c|}{ $a_v$ }&
\multicolumn{1}{|c|}{ $a_s$ }&
\multicolumn{1}{|c|}{ $a_{\rm cv}$ }&
\multicolumn{1}{|c|}{ $a_x$ }&
\multicolumn{1}{|c|}{ $a_y$ }\\
\hline
3 & 865.4 & $-$2.68 & 161.1 & & \\
4 & 865.1 &    1.19 & 145.0 & 18.5 & \\
5 & 865.2 &   0.015 & 152.4 & $-$0.015 & 15.5\\
\hline
3 & 1542.9&  $-$4.0& 233.8& & \\
4 & 1542.5&   1.79 & 209.8&27.7& \\
5 & 1542.6&$-$0.096& 221.7&$-$1.93& 24.9 \\
\hline
3 & 2175.7 & $-$3.56 & 194.9 & & \\
4 & 2175.4 &    1.55 & 173.8 & 24.4& \\
5 & 2175.4 &    0.053& 183.2 & 0.86 & 19.8\\
\hline
\end{tabular}
\end{center}
\end{document}